\documentclass[prl, superscriptaddress, showkeys, reprint, twocolumn, 10pt]{revtex4}
\pdfoutput=1 

\usepackage{graphicx} 
\usepackage{epsfig}
\usepackage{epsf}
\usepackage{epstopdf}
\usepackage{hyperref}
\usepackage{color}

\usepackage{amsmath}  
\usepackage{amsfonts}
\usepackage{amssymb}
\usepackage{pdfsync}
\usepackage{mathrsfs}
\newcommand{\bone}{1\!\!1}
\newcommand{\delC}{\mathscr{G}}
\newcommand{\delCmn}{\mathscr{G}_{\mu\nu}}

\begin{document}
\title{Integral transforms of the quantum mechanical path integral:\\ Hit function and path averaged potential.}
\author{James P. Edwards}
\email[Corresponding author:]{jedwards@ifm.umich.mx}
\affiliation{Instituto de F\'isica y Matem\'aticas,
Universidad Michoacana de San Nicol\'as de Hidalgo,
Edificio C-3, Apdo. \!Postal 2-82,
C.P. 58040, Morelia, Michoac\'an, Mexico}

\author{Urs Gerber}
\email[Email:]{gerber@correo.nucleares.unam.mx}
\affiliation{Instituto de F\'isica y Matem\'aticas,
Universidad Michoacana de San Nicol\'as de Hidalgo,
Edificio C-3, Apdo. \!Postal 2-82,
C.P. 58040, Morelia, Michoac\'an, Mexico}
\affiliation{Instituto de Ciencias Nucleares, Universidad Nacional Aut\'onoma de Mexico, A.P. 70-543, C.P. 04510 Ciudad de M\'{e}xico, Mexico}

\author{Christian Schubert}
\email[Email:]{schubert@ifm.umich.mx}
\affiliation{Instituto de F\'isica y Matem\'aticas,
Universidad Michoacana de San Nicol\'as de Hidalgo,
Edificio C-3, Apdo. \!Postal 2-82,
C.P. 58040, Morelia, Michoac\'an, Mexico}

\author{Maria Anabel Trejo}
\email[Email:]{mtrejo@ifm.umich.mx}
\affiliation{Instituto de F\'isica y Matem\'aticas,
Universidad Michoacana de San Nicol\'as de Hidalgo,
Edificio C-3, Apdo. \!Postal 2-82,
C.P. 58040, Morelia, Michoac\'an, Mexico}
\affiliation{Theoretisch-Physikalisches Institut, Friedrich-Schiller-Universit\"{a}t Jena, Max-Wien-Platz 1, 07743 Jena, Germany}

\author{Axel Weber}
\email[Email:]{axel@ifm.umich.mx}
\affiliation{Instituto de F\'isica y Matem\'aticas,
Universidad Michoacana de San Nicol\'as de Hidalgo,
Edificio C-3, Apdo. \!Postal 2-82,
C.P. 58040, Morelia, Michoac\'an, Mexico}



\begin{abstract}
\noindent We introduce two new integral transforms of the quantum mechanical transition kernel that represent physical information about the path integral. These transforms can be interpreted as probability distributions on particle trajectories measuring respectively the relative contribution to the path integral from paths crossing a given spatial point (the \textit{hit function}) and the likelihood of values of the line integral of the potential along a path in the ensemble (the \textit{path averaged potential}). 
\end{abstract}
\maketitle

\section{Introduction}
\noindent In the standard quantum mechanics of a particle in a time-independent potential $V(\bf r)$, 
a fundamental quantity is the propagator, $K(y, x; T)$, defined as the configuration space matrix element:
\begin{equation}
	K(y, x; T) = \left<y\left|e^{-iHT}\right|x\right>, 
\end{equation}
with Hamiltonian $H= \frac{p^2}{2m} + V({\bf r})$ (using natural units). The propagator holds complete information on the time evolution of the system, satisfying the Schr\"{o}dinger equation, $i\partial_{T} K(y, x; T) = H(y) K(y, x; T)$ with $H(y) = -\frac{1}{2m}\partial^{2}_{y} + V(y)$, so knowing its explicit form is tantamount to solving the system.

In a basis of eigenfunctions of $H$ the propagator has spectral representation (henceforth working in Euclidean space-time so $K(T) = e^{-TH}$)
\begin{align} \sum_{n} \psi_{n}(y)\psi_{n}^{\star}(x)e^{-E_{n} T} + \int dk\,\psi_{k}(y)\psi_{k}^{\star}(x)e^{-E(k)T}
	\label{kSpec}
\end{align}
separated into contributions from bound states and scattering states.

The kernel for a scalar particle also has path integral representation
\begin{align}
	K(y, x; T) &= \int_{x(0) = x}^{x(T) = y} \hspace{-1.5em} \mathscr{D}x \,e^{-\int_{0}^{T} dt\left[ \frac{m\dot{x}^{2}}{2} + V(x(t)) \right] }
	\label{Kpi}
\end{align}
with free particle normalisation 
\begin{equation}
	K_{0}(y, x; T) =\int_{x(0) = x}^{x(T) = y} \hspace{-1.5em} \mathscr{D}x \,e^{-\int_{0}^{T}dt\,	 \frac{m\dot{x}^{2}}{2}} = \left(\frac{m}{2\pi T}\right)^{\frac{D}{2}} e^{-\frac{m\left(y-x\right)^{2}}{2T}}.
	\label{K0}
\end{equation}

This paper introduces two novel representations of the propagator: the ``hit function'' $\mathcal{\overline{H}}(z | y, x; T)$ and the ``path averaged potential'' $\mathcal{\overline{P}}(v | y, x; T)$. Both are invertible integral transforms of the kernel,	
\begin{align}
	    K(y, x; T) &= \int d^{D}z\, \mathcal{\overline{H}}(z| y, x; T), \label{KHz}  \\
		K(y, x; T) &= \int_{-\infty}^{\infty} \!dv\, \overline{\mathcal{P}}(v | y, x; T) e^{-v}.	
		\label{KPv}
\end{align} 
Our original motivation for these representations lies in their usefulness for numerical sampling of the path integral \eqref{Kpi} and we report our
corresponding results elsewhere \cite{UsReg}. 
However, we anticipate that these representations will find much wider application,  providing specific physical information on the dynamics of the system. 
The hit function has an analogy in the proper time formalism of quantum field theory (see \cite{polybook, ChrisRev}), whilst a relativistic version of the path averaged potential was introduced by Gies et al. \!\cite{gisava}; we adapt these objects for quantum mechanics, supplying the inverse transformations lacking in \cite{polybook} and \cite{gisava} and their gauge dependence.

We first define $\mathcal{\overline{P}}(v)$ and $\mathcal{\overline{H}}(z)$ and provide the inverse transformations to (\ref{KHz}) and (\ref{KPv}). We then discuss their asymptotic form for large $T$ and their gauge transformations for electromagnetic interactions. Finally we calculate both functions for some simple potentials and show compatibility with numerical samplings.

\section{Transforms of the kernel}
\label{secTrans}
\noindent In theoretical calculations and numerical simulations a direct determination of the kernel can be difficult, so it may be advantageous to consider some intermediate quantity. Here we present two such quantities.
\subsection{The hit function, $\mathcal{\overline{H}}(z)$}
\noindent The first quantity is a function of a spatial point, $z$, measuring the relative contribution to the propagator of paths which pass through $z$. This object, $\mathcal{\overline{H}}(z)$, which we call the ``hit function'' is defined by
\begin{align}
	&\mathcal{\overline{H}}(z | y, x; T) \equiv  \frac{1}{T}\int_{x(0) = x}^{x(T) = y} \hspace{-1.8em}\mathscr{D}x \int_{0}^{T}\! \delta^{D}\left(z -  x(\tau)  \right) d\tau\, e^{-S[x]},
	\label{Hz}
\end{align}
where $S[x] = \int_{0}^{T}dt\left[ \frac{m\dot{x}^{2}}{2} + V(x(t))\right]$ is the (Euclidean) classical action of the trajectory $x(t)$. The $\delta$ function counts the worldlines that cross through, or hit, the point $z$, weighting each path by its action. $\mathcal{\overline{H}}(z)$ is inspired by a similar quantity called ``local time'' \cite{Levy}, measuring the time a fixed trajectory spends at a given point \cite{localtime2, localtime1}. We have generalised this to include the weight associated to each trajectory and the external potential in computing its first moment. The path integral (\ref{Hz}) is also similar to the contact interactions used in \cite{MeCont, MeStr} and the field theory amplitudes developed in \cite{polybook}. For later calculations it is convenient to introduce a properly normalised distribution on the space of trajectories
\begin{equation}
	\mathcal{H}(z | y, x; T) \equiv \frac{\mathcal{\overline{H}}(z | y, x; T)}{K(y, x; T)},
	\label{Hnorm}
\end{equation}
which integrates to unity.

Since $\mathcal{\overline{H}}(z)$ is built only out of particle worldlines that pass through the point $z$ we can relate it to the kernel by forcing this criterion at some arbitrary time $0 < t < T$, leading to an inverse integral transform to (\ref{KHz}),
\begin{equation}
	\mathcal{\overline{H}}(z | y, x; T) = \frac{1}{T}\int_{0}^{T} dt\, K(z, x; t)\, K(y, z; T - t).
	\label{Hkern}
\end{equation}
Note that the kernels in the integrand count \textit{all} paths between the end-points, including those that cross through $z$ multiple times, in agreement with the $\delta$ function in (\ref{Hz}). The Feynman Green function has been similarly decomposed in terms of restricted kernels in the ``path decomposition expansion'' of \cite{PDX3, PDX1, PDX2}.
%

There are many approaches to numerical evaluation of the quantum mechanical path integral, such as Monte-Carlo sampling \cite{FeynKac1, MonteStat, FeynKac2, MonteRev, Makri, Olof} or, as we employ in \cite{UsReg}, the adaptation of worldline numerics \cite{Tjon, GiesMagnetic, GiesClouds, GiesCasimir} to the non-relativistic setting. In such cases, where the goal may be estimation of the kernel, knowledge of the hit function is sufficient, for one need simply integrate over positions in (\ref{Hz}) or (\ref{Hkern}) to verify (\ref{KHz}). So one could sample the hit function via its path integral representation (\ref{Hz}) and determine the kernel through finite dimensional (numerical) integration.

\subsection{The path averaged potential, $\mathcal{\overline{P}}(v)$}
\noindent Path integral determination of the kernel amounts to the expectation value of exponentiated line integrals of the potential. This motivates describing the likelihood of values of these line integrals over particle paths with Gaussian distribution on their velocities. Denoting this function by $\mathcal{\overline{P}}(v)$ where $v \equiv \int_{0}^{T} V(x(t)) dt$ we express this likelihood as a constrained path integral:
\begin{align}
	&\mathcal{\overline{P}}(v| y, x; T) \equiv \int_{x(0) = x}^{x(T) = y} \hspace{-1.8em}\mathscr{D}x \,\delta\left(v - \int_{0}^{T} V(x(t)) dt \right) e^{-\int_{0}^{T} \frac{m\dot{x}^{2}}{2} dt} .
	\label{PvPI}
\end{align}
Prior use of a similar function has been made in relativistic quantum theory \cite{gisava}, where worldline numerics are used to sample the likelihood (a function of proper time) and functional fits are made to the numerical results.

Using the Fourier representation of the $\delta$ function we may write $\mathcal{\overline{P}}(v)$ in terms of the kernel, 
\begin{align}
\mathcal{\overline{P}}(v | y, x; T) &=\frac{1}{2\pi}\int_{-\infty}^{\infty} dz\, e^{ivz} \tilde{K}(y, x; T,z),
	\label{PvK}
\end{align}
where $\tilde{K}$ relates to $K$ by the substitution $V(x) \rightarrow izV(x)$ under the path integral. This follows if one may exchange the functional and Fourier integrations and gives the inverse transform to (\ref{KPv}).

Again we also introduce a normalised probability distribution by
\begin{equation}
	\mathcal{P}(v | y, x; T) \equiv \frac{\mathcal{\overline{P}}(v | y, x; T)}{K_{0}(y, x; T)},
	\label{Pnorm}
\end{equation}
which has unit area. Note in contrast to (\ref{Hnorm}) the normalisation for $\mathcal{P}(v)$ is always known analytically since it involves only the free kernel.  

It is straightforward to check that  (\ref{KPv}) follows from (\ref{PvPI}) or (\ref{PvK}), so numerical estimation of the kernel follows from the expectation of $e^{-v}$ against the likelihood $\mathcal{\overline{P}}(v)$, again reducing the problem to a finite dimensional integral.

\subsection{Asymptotic properties of the distribution functions}
\noindent The spectral representation of the kernel (\ref{kSpec}) implies asymptotic formulae for our new functions as $T  \to \infty$, relevant when extracting the ground state energy, for example. We focus on the normalised distributions defined in (\ref{Hnorm}) and (\ref{Pnorm}). 

For $\mathcal{H}(z)$ we substitute the spectral decomposition directly into (\ref{Hkern}). Subsequently integrating over the intermediate time $t$ leads to a double sum
\begin{widetext}
\begin{equation}
	\mathcal{H}(z | y, x; T) = \frac{1}{T K(y, x; T)} \left[T \sum_{n}\psi_{n}(y)\left|\psi_{n}(z)\right|^{2}\psi_{n}^{\star}(x) e^{-E_{n}T} + \sum_{n, m \neq n} \psi_{n}(y) \psi_{n}^{\star}(z)\psi_{m}(z) \psi_{m}^{\star}(x) \frac{e^{-E_{n}T} - e^{-E_{m}T}}{E_{m} - E_{n}} \right],
	\label{Hasymt}
\end{equation}
\end{widetext}
where we have only included contributions from bound states; indeed as $T \rightarrow \infty$, the leading contributions arise from states with least energy. Denoting the ground state energy (assumed non-degenerate) by $E_{0}$ we find the asymptotic form
\begin{align}
	&\mathcal{H}(z | y, x; T) \simeq \left|\psi_{0}(z)\right|^{2} + \nonumber \\
	&\frac{1}{T \psi_{0}(y)\psi_{0}^{\star}(x)} \sum_{n > 0} \frac{\psi_{0}(y)\psi_{0}^{\star}(z)\psi_{n}(z)\psi_{n}^{\star}(x) + (n\leftrightarrow 0) }{E_{n} - E_{0}}
	\label{Hlo}
\end{align}
up to contributions of order $e^{-(E_{1} - E_{0})T}$. Note the leading order term on the right hand side has no dependence upon initial and final positions, as expected for late time evolution.

For $\mathcal{P}(v)$, sending $V(x) \rightarrow izV(x)$ modifies the Hamiltonian. Calculation of $\mathcal{P}(v)$ supposes the kernel can be analytically continued. If the spectral decomposition can also be continued, with new ``wavefunctions,'' $\widetilde{\psi}(x;z)$, and ``energies'' $\widetilde{E}_{n}(z)$ (we shall see that for the harmonic oscillator with frequency $\omega$, $E_{n} \rightarrow \widetilde{E}_{n}(z) = \frac{1 \pm i}{\sqrt{2}}\omega \sqrt{|z|} (n + \frac{1}{2})$) whose real parts remain bounded, there will still exist an energy, $\widetilde{E}_{0}$, satisfying $\Re(\widetilde{E}_{0}) \leqslant \Re(\widetilde{E}_{n})$ for $n > 0$. Then we get a large-$T$ asymptotic formula
\begin{equation}
	\widetilde{K}(y, x; T) \simeq \sum_{n_{0}} \widetilde{\psi}_{n_{0}}(y; z)\widetilde{\psi}^{\star}_{n_{0}}(x;z)e^{-\widetilde{E}_{n_{0}}(z) T}
	\label{Ktildasyp}
\end{equation}
where the sum is over states with $\Re(\widetilde{E}_{n_{0}}) = \Re(\widetilde{E}_{0})$. Substituting this asymptotic representation into (\ref{PvK}) we learn that for large $T$,
\begin{align}
	&\mathcal{P}(v | y, x; T) \simeq \nonumber \\
	&\frac{1}{2\pi K_{0}(y, x; T)}\sum_{n_{0}}\int_{-\infty}^{\infty} dz e^{ivz}\widetilde{\psi}_{n_{0}}(y; z)\widetilde{\psi}^{\star}_{n_{0}}(x;z)e^{-\widetilde{E}_{n_{0}}(z) T},
\end{align}	
giving one approach to studying the asymptotics of $\mathcal{P}(v)$. 

\subsection{Electromagnetic interactions}
\noindent For a particle coupled to a gauge potential, $\mathcal{A}$, we must also consider the inherent gauge freedom. The (Euclidean) action is 
\begin{equation}
	S[x, \mathcal{A}] = \int_{0}^{T}dt\,\left[ \frac{m\dot{x}^{2}}{2} + ie\mathcal{A}(x(t))\cdot \dot{x}\right].
\end{equation}
Given a reference gauge, with potential $\mathcal{\hat{A}}$, the kernel with respect to this gauge changes covariantly under a gauge transformation  $\mathcal{\hat{A}}_{\mu}(x) \rightarrow \mathcal{A}_{\mu}(x) = \mathcal{\hat{A}}_{\mu}(x) + \partial_{\mu}\Lambda(x)$ as \cite{HolgDittr}
\begin{align}
	K(y, x; T) &= e^{-ie\varphi(y,x)}\hat{K}(y, x; T),
	\label{Kgt} 
\end{align}
with $\varphi(y,x) \equiv \int_{x}^{y} (\mathcal{A} - \mathcal{\hat{A}})\cdot dx = \Lambda(y) - \Lambda(x)$.

The distribution $\mathcal{H}(z)$ is gauge invariant, although it now becomes complex valued. Using formula (\ref{Hkern}) in a specific gauge,
\begin{equation}
	\mathcal{\hat{H}}(z | y, x; T) = \frac{1}{T \hat{K}(y, x; T)} \int_{0}^{T}\! dt\, \hat{K}(z, x; t)\hat{K}(y, z; T-t),
\end{equation}
gauge transforming the kernels  leads to the transformed hit function distribution
\begin{equation}
	\mathcal{H}(z | y, x; T) = \frac{e^{-ie (\varphi(z,x) + \varphi(y,z))}}{ e^{-ie \varphi(y,x)}}\hat{\mathcal{H}}(z | y, x; T).
\end{equation}
Here we have used the fact that the holonomy is independent of the path. The phase factors cancel so that $\mathcal{H}(z | y, x; T) = \mathcal{\hat{H}}(z | y, x; T)$ ($\overline{\mathcal{H}}(z)$ picks up a phase $-ie \varphi(y, x)$) and we may choose a convenient gauge for its computation.

We also modify our formula for $\mathcal{P}(v)$ to become a real-valued distribution on the phase introduced by the potential. This distribution will be gauge dependent, so we define $\mathcal{\hat{P}}(v | y, x; T)$ with respect to our reference gauge as
\begin{align}
	&\frac{1}{{K_{0}(y, x; T)}}\int_{x(0) = x}^{x(T) = y} \hspace{-1em}\mathscr{D}x\, \delta\left(v - e\int_{0}^{T} \mathcal{\hat{A}}\cdot \dot{x} \,d\tau \right) e^{-\int_{0}^{T} dt \frac{m\dot{x}^{2}}{2}} \nonumber \\
	=&\frac{1}{2\pi{K_{0}(y, x; T)}} \int_{-\infty}^{\infty} dz \,e^{i v z} \widetilde{\hat{K}}(y, x; T, z),
	\label{Pvkgauge}
\end{align}
where now $\widetilde{\hat{K}}$ requires scaling $e \rightarrow ze$. Under a gauge transformation to a general potential $\mathcal{A}$ the distribution becomes
\begin{align}
&\frac{1}{2\pi K_{0}(y, x; T)}\int_{-\infty}^{\infty}dz e^{ivz -ie \varphi(y,x) z}\widetilde{\hat{K}}(y, x; T,z) \nonumber \\
	\label{Pgauge}
\end{align}
which is just $\mathcal{\hat{P}}(v - e \varphi(y,x) | y, x; T)$, so changing gauge translates the distribution by $e (\Lambda(y) - \Lambda(x))$ (the same holds for $\overline{\mathcal{P}}(v)$). These gauge transformations follow also from the definitions (\ref{Hz}) and (\ref{PvPI}).

\section{\label{secApp}Applications}
\noindent In this section we give explicit expressions for the distributions $\mathcal{H}(z)$ and $\mathcal{P}(v)$ for some simple potentials and compare our analytic results to sampling based upon worldline numerics. We cover quadratic potentials followed by a constant magnetic field, calculating the distributions using standard path integral techniques. We present detailed results on numerical estimation of their kernels elsewhere \cite{UsReg}.

\subsection{The hit function}
\begin{figure}
	\hspace{-1em}\includegraphics[width=0.5\textwidth]{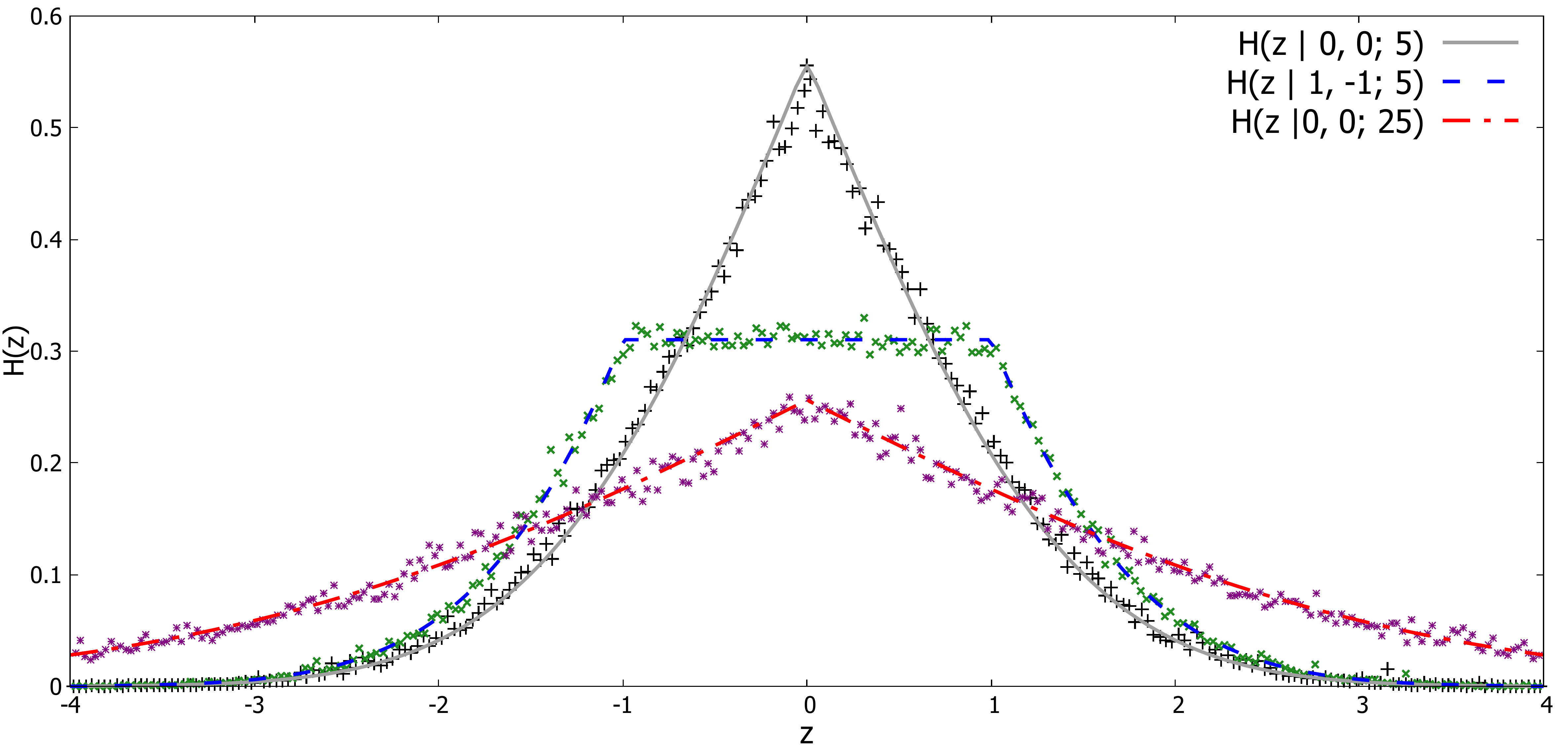}
	\caption{$\mathcal{H}(z | y, x; T)$, for the free particle in one dimension. The solid line plots (\ref{HfreeErf}) and the data points represent a numerical sampling. Translating the endpoints shifts the hit function due to translational symmetry.}
	\label{figHfree}
\end{figure}

\noindent For $\mathcal{H}(z)$, it is instructive to use (\ref{Hz}). For illustration, consider an action quadratic in the trajectory ($m = 1$ henceforth),
\begin{equation}
	S[x] = \int_{0}^{T} dt\left[x \cdot \frac{1}{2}M \cdot x + b \cdot x \right],
\end{equation}
whose classical solution satisfies $M \cdot x_{c} + b = 0$ with boundary conditions $x_{c}(0) = x$, $x_{c}(T) = y$. Expanding about this solution and invoking the Fourier representation of the $\delta$ function, (\ref{Hnorm}) and (\ref{Hz}) yield
\begin{align}
	T\mathcal{H}(z | y, x; T) = \int_{0}^{T}d\tau \int \frac{d^{D} s}{(2\pi)^{D}}\, e^{is \left(z - x_{c}(\tau)\right) - \frac{s^{2}}{2} G_{M}(\tau, \tau)},
\end{align}
where $G_{M}(\tau, \tau^{\prime})$ is the ``worldline'' Green function \cite{ChrisRev, 25} for $M$ satisfying Dirichlet boundary conditions $G_{M}(0, \tau^{\prime}) = 0 = G_{M}(T, \tau^{\prime})$. The Gaussian integral over $s$ provides
\begin{equation}
	\mathcal{H}(z | y, x; T) = \int_{0}^{T} \frac{d\tau}{T \left(2 \pi G_{M}(\tau, \tau)\right)^{\frac{D}{2}}} e^{-\frac{\left(z - x_{c}(\tau)\right)^{2}}{2G_{M}(\tau, \tau)}}
	\label{Hquad}
\end{equation}
which is suitable for numerical integration. For the free particle the Green function for $M = -\frac{d^{2}}{dt^{2}}$ is $G_{M}(t, t^{\prime}) = -\Delta(t, t^{\prime})$ where $\Delta(t, t^{\prime}) = \frac{1}{2}\left|t - t^{\prime}\right| - \frac{1}{2}(t + t^{\prime}) + \frac{ t t^{\prime}}{T}$, and $x_{c}(t)$ is the straight line path from $x$ to $y$, so that in one dimension the hit function is
\begin{equation}
	\mathcal{H}_{0}(z | y, x; T) = \int_{0}^{1} \frac{du}{\sqrt{2 \pi T u (1 - u)}}e^{-\frac{\left(z - x - (y-x)u\right)^{2}}{2T u(1 - u)}}.
	\label{Hfree}
\end{equation}
Here we have scaled $\tau = Tu$ and used $x_{c}(t) = x + (y - x)\frac{t}{T}$. This also follows from (\ref{Hkern}), using (\ref{K0}), to verify the inversion formula. The integral can be written in terms of the complementary error function\footnote{We are indebted to V\'{a}clav Zatloukal for indicating local-time calculations relevant for the free particle. We use the standard notation $\textrm{Erfc}(\chi) \equiv 1 - \frac{2}{\sqrt{\pi}} \int_{0}^{\chi} e^{-\rho^{2}}d\rho$.} as
\begin{equation}
	\mathcal{H}_{0}(z | y, x; T) = \sqrt{ \frac{\pi}{2T} } e^{ \frac{(x - y)^2}{2T} } \textrm{Erfc}\left[ \frac{|z - x|  + |z - y|} { \sqrt{2T}} \right] ,\label{HfreeErf}
\end{equation}
familiar in the context of L\'{e}vy random walks \citep{localtime1}. In figure \ref{figHfree} we show the distribution for suitable values of $x$, $y$ and $T$ along with simulated samples based on worldline numerics. 

\begin{figure}
	\hspace{-1em}\includegraphics[width=0.5\textwidth]{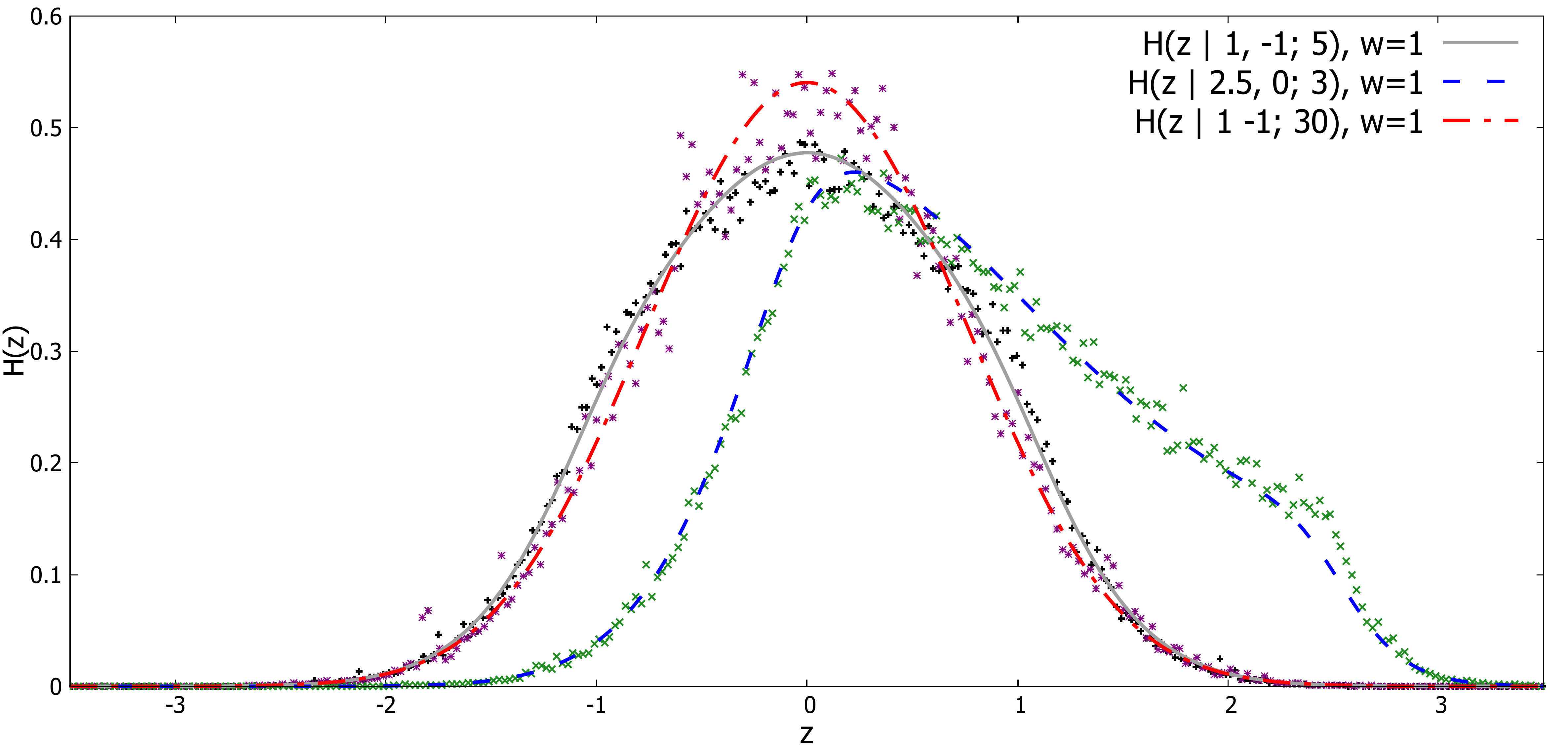}
	\caption{$\mathcal{H}(z | y, x; T)$ for the harmonic oscillator in one dimension, $\omega = 1$. The solid line is a numerical evaluation of the hit function distribution and the data points represent a numerical sampling; increasing $T$, the distribution resembles $\left|\psi_{0}(z)\right|^{2}$ as in (\ref{Hlo}).}
	\label{figHHO}
\end{figure}

For the one dimensional linear potential ($M = -\frac{d^{2}}{dt^{2}}$, $b = k$) the Green function is unchanged but the classical solution becomes $x_{L}(t) = x + \left(\frac{y - x}{T} - \frac{kT}{2} \right)t + \frac{kt^{2}}{2}$ so that (\ref{Hquad}) must be integrated numerically. Worldline numerics are in  excellent agreement with the result that ensues. For the harmonic oscillator, $M = -\frac{d^{2}}{dt^{2}} + \omega^{2}$, and we require the coincident Green function \cite{kleinbook}
\begin{align}
	G_{\omega}(\tau, \tau) &= \frac{1}{\omega}\frac{\sinh(\omega \tau)\sinh(\omega (T - \tau))}{\sinh(\omega T)}.
\end{align}
We show the numerical evaluation of (\ref{Hquad}) for the harmonic oscillator and its correspondence with a sampling of the distribution using worldline numerics in figure \ref{figHHO}. One may verify the formulae via (\ref{Hkern}) using the well known kernels given in \cite{Groesche}.

Note that for actions that are not quadratic and where it is not feasible to compute the path integral, formula (\ref{Hquad}) can be applied as a semi-classical approximation given sufficiently good knowledge of $x_{c}$.


For particle motion in a plane threaded by a perpendicular, constant magnetic field, $B$, Fock-Schwinger gauge about the endpoint $x$ reduces the action to one that is quadratic in the trajectory \cite{scalar} so that we may use (\ref{Hquad}) with $M = -\frac{d^{2}}{dt^{2}}\bone + ieF \cdot \frac{d}{dt}$. The coupling to the gauge potential is absorbed into the worldline Green function, $\delCmn(t, t^{\prime})$ given in the appendix. We expand about the straight line path, $x_{0}(t)$, so that
\begin{equation}
	\mathcal{H}(z | y, x; T) = \int_{0}^{T}\frac{dt}{2\pi T}\frac{ e^{-\frac{1}{2}\left( z - x(t)\right)^{\intercal} \cdot \delC^{-1}(t, t) \cdot \left( z - x(t)\right)}}{\textrm{det}[\delC(t, t)]}
\end{equation}
where $x(t) = x_{0}(t) - \frac{e}{T}\int_{0}^{T} \delC(t, t^{\prime})dt^{\prime} \cdot (y-x)$.
\subsection{The path averaged potential}
\begin{figure}
	\hspace{-1em}\includegraphics[width=0.5 \textwidth]{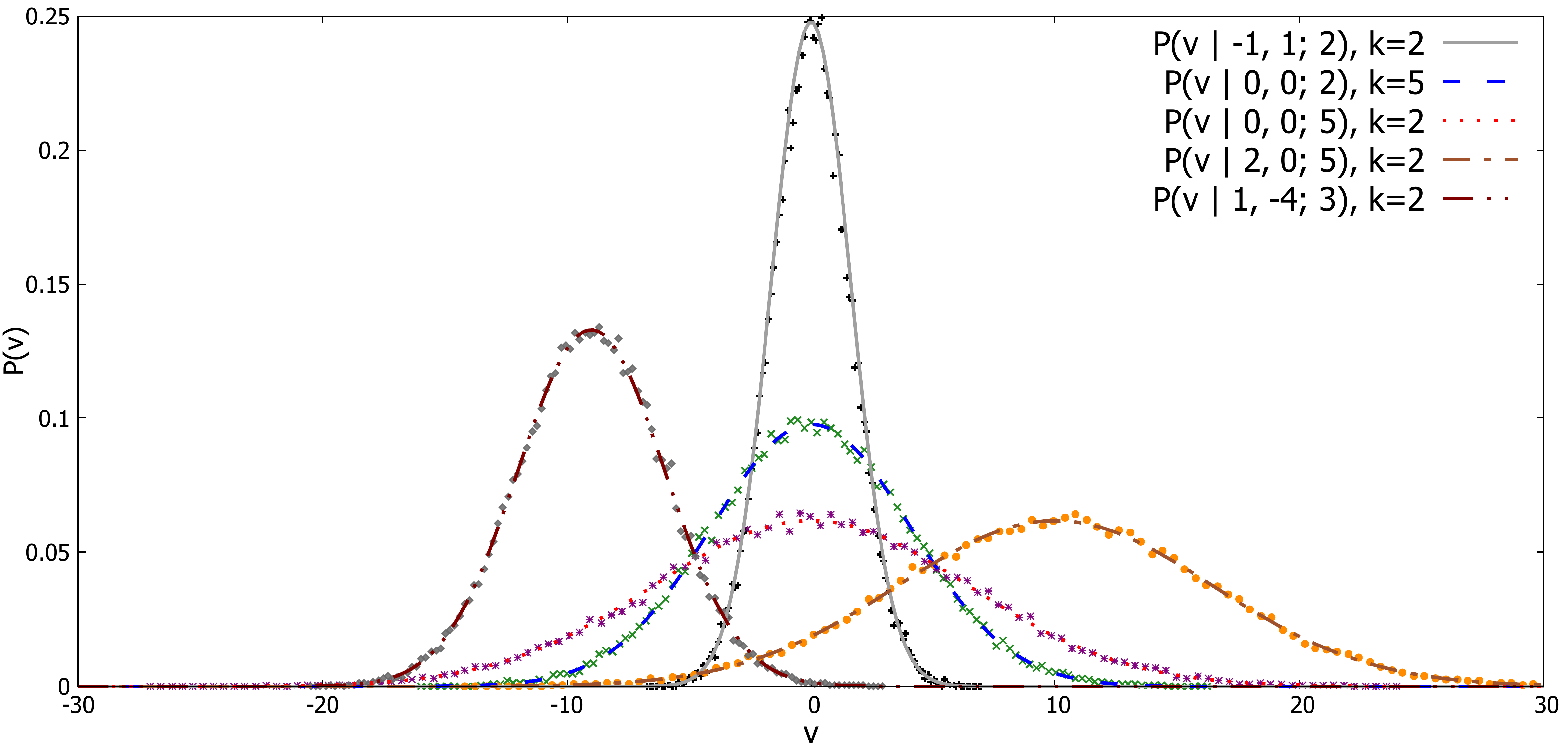}
	\caption{$\mathcal{P}(v)$ for the linear potential in one dimension. The points represent a computational sampling of the distribution using worldline numerics; lines are theoretical fits based upon (\ref{PvLin}).}
	\label{figPvLin}
\end{figure}

\noindent Turning to the path averaged potential, (\ref{PvPI}) shows that for a free particle $\mathcal{P}(v) = \delta(v)$. For the linear potential, $V(x) = kx$, the kernel is \cite{Groesche}
\begin{align}
	& K(y, x; T) = \sqrt{\frac{1}{2\pi T}}\, e^{\left[-\frac{1}{2T}\left(x-y\right)^{2} -\frac{kT}{2}(x+y)+ \frac{k^{2}T^{3}}{24}\right]},
\end{align}
and to effect the change $V \rightarrow izV$ it suffices to send $k \rightarrow izk$. Application of (\ref{PvK}) supplies
\begin{equation}
	\mathcal{P}(v | y, x; T) = \sqrt{\frac{6}{\pi k^{2}T^{3}}}e^{-\frac{3}{2T}(x + y)^{2}}e^{-\frac{6}{k^{2}T^{3}}\left(v^{2} - kT(x +y)v\right)},
	\label{PvLin}
\end{equation}
which is a function of the sum of the endpoints. In figure \ref{figPvLin} we demonstrate agreement with numerical sampling. The limit $k \rightarrow 0$ supplies the $\delta(v)$ of the free particle and one may verify (\ref{KPv}) from (\ref{PvLin}) directly.

For the harmonic oscillator, (\ref{PvK}) leads to a highly oscillatory $z$-integral that must be evaluated numerically. Since the spectrum consists only of bound states, we apply instead the scaling $\omega \rightarrow \sqrt{iz}\omega$ directly in the spectral decomposition and take the real part of the $z$-integral along the positive real line. Then the energies scale to $\widetilde{E}_{n} = \frac{1 + i}{2}\sqrt{z} E_{n}$. This leads to a sum 
\begin{align}
	&\pi K_{0}(y, x; T)\mathcal{P}(v | y, x; T) =  \sqrt{\frac{\omega}{\pi}}\sum_{n \geqslant 0} \frac{1}{2^{n} n!} \times \nonumber \\
	&\Re\int_{0}^{\infty} dz (iz)^{\frac{1}{4}}e^{ivz} e^{-\frac{1}{2}(\tilde{x}^{2} + \tilde{y}^{2}) - \sqrt{iz}\omega T (n + \frac{1}{2})}H_{n}(\tilde{x})H_{n}(\tilde{y})
\end{align}
for $\tilde{x}^{2} \equiv  \sqrt{iz}\omega x^{2}$ and $\tilde{y}^{2} \equiv \sqrt{iz}\omega y^{2}$. To make analytic headway, we set $x = 0 = y$; the $z$-integral can now be written in terms of modified Bessel functions of the second kind, $K_{n}$. The resulting sum (setting $v_{n} = \frac{((n+\frac{1}{2})\omega T)^{2}}{8v}$ for brevity),
\begin{align}
\mathcal{P}(v | 0, 0; T) &= 64\Theta(v)\sqrt{\frac{\omega T}{2 \pi^{2}}} \sum_{n \textrm{ even}} \frac{n! \, v_{n}^{\frac{3}{2}} e^{-v_{n}} }{2^{n+\frac{1}{2}} (\frac{n}{2})!^{2} \left((n+\frac{1}{2}) \omega T \right)^{\frac{5}{2}}}\nonumber \\
	&\times \Re \Bigg[\left(v_{n} - \frac{3}{4} \right)K_{-\frac{1}{4}}\left(-v_{n}\right) - v_{n} K_{-\frac{5}{4}}\left(-v_{n}\right)\Bigg]
\end{align}
converges extremely rapidly so is apt for truncation to arbitrary accuracy ($\Theta(v)$ is the heaviside step function). We have checked this gives excellent agreement with sampled data generated by worldline numerics (truncation to $n \leqslant 30$ is more than sufficient) which will be presented in \cite{UsReg}.

\begin{figure}
	\centering
	\hspace{-1em}\includegraphics[width=0.5\textwidth]{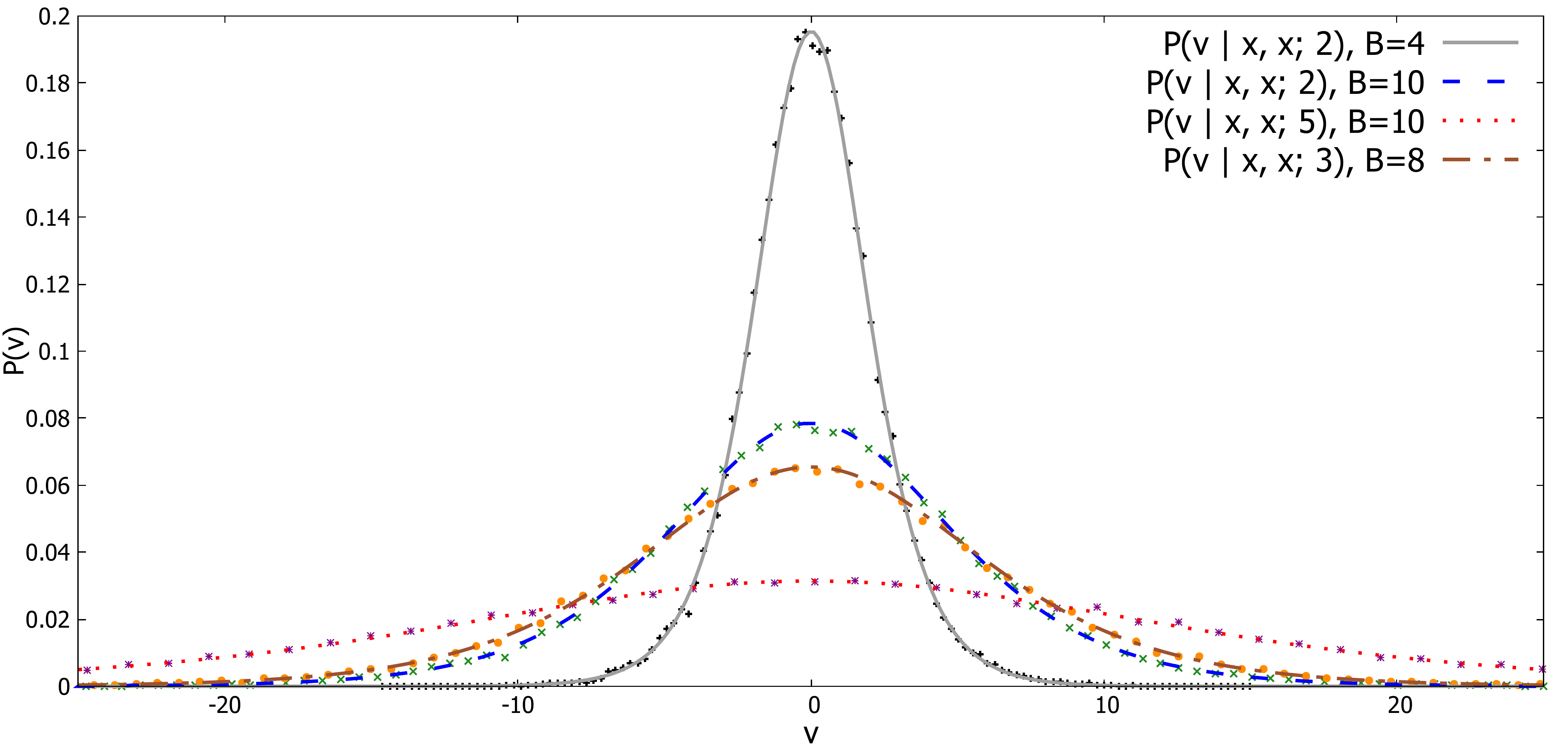}
	\caption{$\mathcal{\hat{P}}(v)$ for a constant magnetic field (Fock-Schwinger gauge) and $x = y$ for illustrative values of $T$ and $B$. Lines are analytic curves using (\ref{PvB}) overlayed on a sampling of the distribution using worldline numerics.}
	\label{figPvB}
\end{figure}

Finally we return to the case of a constant magnetic field. In Fock-Schwinger gauge, the reference kernel evaluates to \cite{Groesche}
\begin{equation}
	\hat{K}(y, x; T) = \frac{eB}{4 \pi \sinh(\frac{e B T}{2})}e^{-\frac{eB}{4}|x - y|^{2} \coth(\frac{e B T}{2})}.
	\label{KhatB}
\end{equation}
Sending $e \rightarrow ze$ and using (\ref{Pvkgauge}) we must evaluate
\begin{equation}
	\frac{eBT}{4\pi}\int_{-\infty}^{\infty} dz\, \frac{z}{\sinh(\frac{eBTz}{2})}e^{ivz -\frac{eBz}{4}|x-y|^{2} \coth(\frac{eBTz}{2}) }.
\end{equation}
This is easily evaluated numerically but we can make analytic progress for the diagonal elements, which by translational symmetry are all equal. The $z$-integral can be computed by closing the contour in the upper half plane, leading to
\begin{equation}
	\mathcal{\hat{P}}(v | x, x; T) = \frac{\pi}{2 e B T} \textrm{sech}^{2}\left( \frac{\pi v}{e B T} \right).
	\label{PvB}
\end{equation}
This result bears close similarity to the distributions used in a relativistic setting in \cite{holg3}. See figure \ref{figPvB} for an illustration of the close match provided by worldline numerics. By taking $B\rightarrow 0$ we acquire a representation of the $\delta$ function and (\ref{KPv}) holds after changing $e^{-v} \rightarrow e^{-iv}$. 

\section{Conclusion}
\noindent We have introduced two new integral transforms of the quantum mechanical kernel as tools to study the path integral. These functions contain statistical information about contributions to the path integral of different trajectories. We have given asymptotic formulae and discussed their behaviour under gauge transformations, before demonstrating the distributions with some examples. These calculations are verified by numerical sampling based upon worldline numerics. Elsewhere we shall provide application to more complex systems as an alternative to traditional kernel-based methods. An outstanding issue remains the general validity of the complex continuation of the spectral decomposition to determine $\mathcal{\overline{P}}(v)$. Although we did not rely solely upon this continuation, we recognise that this requires greater scrutiny (as discussed in \cite{PseudSpec, SpecAdj}). We also aim to incorporate spin degrees of freedom into these calculations in future work. 

\subsection*{Acknowledgements}
\begin{acknowledgments}
\noindent JPE and CS wish to thank Petr Jizba and V\'{a}clav Zatloukal for insightful discussions and for sharing their expertise on local time; their thanks also to the organisers of the Path Integration in Complex Dynamical Systems 2017 workshop and the warm hospitality of the Leiden Center where this work was stimulated. MAT thanks Holger Gies and the TPI, Jena, for hospitality and support during the development of the computer code for worldline numerics employed in this analysis. CS would like to thank Erhard Seiler for useful discussions and correspondence. JPE appreciates \'{I}\~{n}igo L. Egusquiza making thoughtful comments and pointing out useful literature related to the hit function. JPE and CS gratefully acknowledge funding from CONACYT through grant Ciencias Basicas 2014 No. 242461 and UG and AW acknowledge funding by Conacyt project no.\ CB-2013/222812. AW is also grateful to CIC-UMSNH for support. 
\end{acknowledgments}
\bigskip

\bibliography{bibHP}

\appendix

\section{Worldline Green function for constant magnetic background}
\noindent The worldline Green function for a particle in a plane with a constant, perpendicular magnetic background field, $B$, satisfying Dirichlet boundary conditions, is \cite{mckshe}:
\begin{align}
\hspace{-1em}\delCmn(t, t^{\prime})& = 	 -\frac{2}{B}\left[\delta_{\mu\nu}\cosh\left(\frac{Bt_{-}}{2}\right) - i\epsilon_{\mu\nu}\sinh\left(\frac{Bt_{-}}{2}\right)\right]\nonumber \\
	\hspace{-1em}   &\times \left[\Theta(t_{-})\sinh\left(\frac{Bt_{-}}{2}\right) - \frac{\sinh\left(\frac{Bt}{2}\right)\sinh\left(\frac{B}{2}(T - t^{\prime}) \right)}{\sinh\left(\frac{BT}{2}\right)}\right]
\end{align}
where $t_{-} \equiv t - t^{\prime}$, $\epsilon_{12} = 1 = -\epsilon_{21}$. 
\end{document}